\documentclass[a4paper]{article}
\usepackage[english]{babel}
\usepackage[T1]{fontenc}
\usepackage[utf8]{inputenc}
\usepackage{graphicx,enumerate}
\usepackage{color}
\usepackage{amsmath, amsfonts, amssymb, mathrsfs, mathdots}
\usepackage{array, booktabs}
\usepackage{ctable, dashrule}

\newcommand{\Lagr}{\mathcal L}
\newcommand{\ddroit}{\mathrm d}

\newcommand{\dt}{\, \mathrm dt}
\newcommand{\matrice}{\mathbf}

\newcommand{\second}{2\textsuperscript{nd}}

\title{Covariant formulation for the optimal control of jointed arm robots:
an alternative to Pontryagin's principle}
\author{J. A. Rojas Quintero, C. Vallée, J. P. Gazeau, P. Seguin}
\date{}

\graphicspath{{./}}

\begin{document}

\maketitle

\textit{Université de Poitiers, Institut Pprime, UPR CNRS 3346, SP2MI, Téléport 2. 11, Boulevard M. et P. Curie, BP 30179, 86962 FUTUROSCOPE-CHASSENEUIL Cedex (France)}

\begin{abstract}
    \textit{We elaborate algorithms able to efficiently command the actuators of an articulated robot. Our time discretization method is based on cubic and quintic Hermite Finite Elements. The suggested control optimization consists in minimizing directly the selected criterium by a conjugate gradient type algorithm. A generic example illustrates the super convergence of the Hermite's technique.}
\end{abstract}


\section{Introduction}

\begin{figure}[htbp]
	\centering
	\def\svgwidth{4cm}
	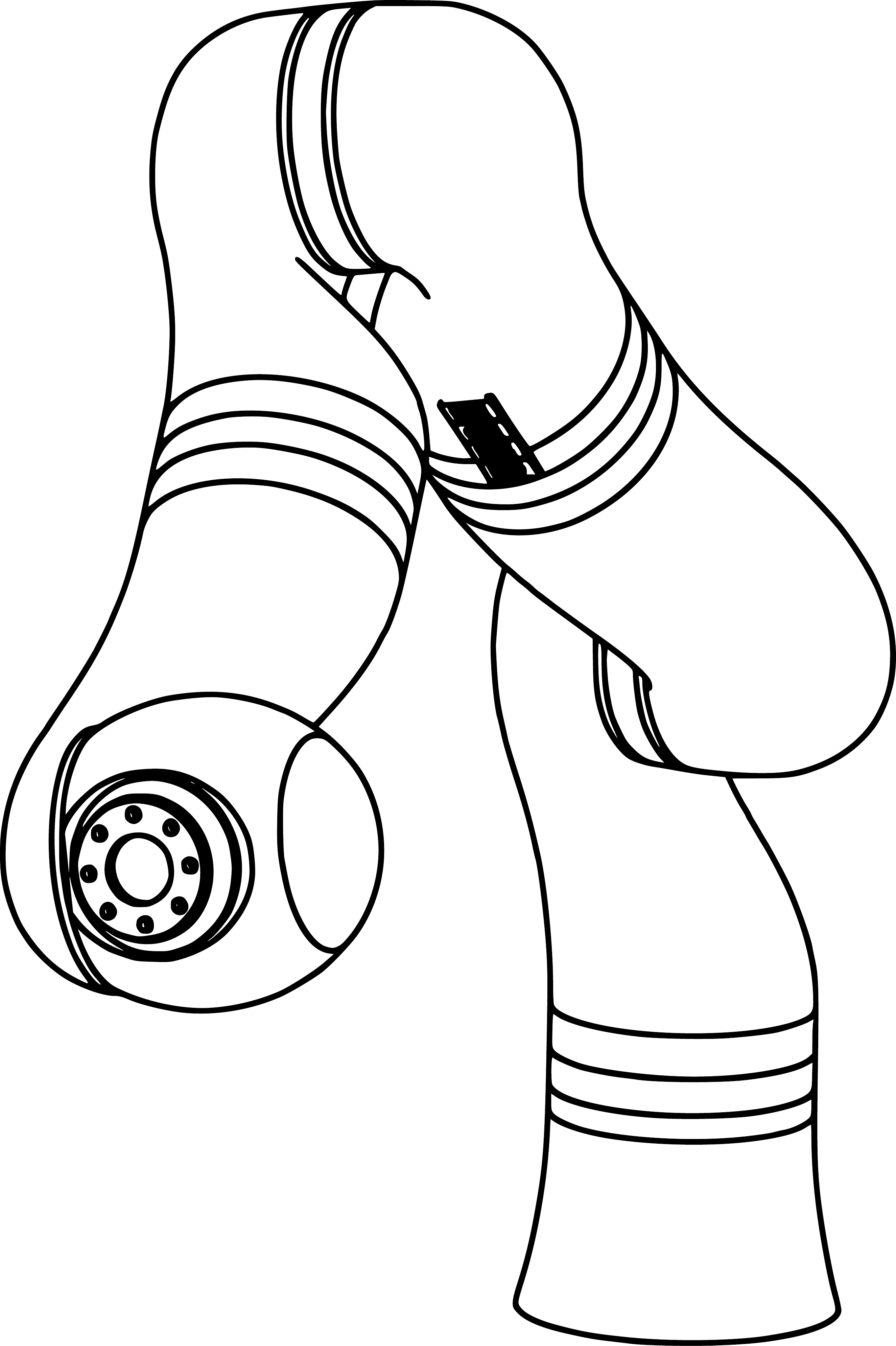
\caption{Robotic manipulator --- \textsc{Kuka LBR}}
\label{fig:kukaLBR4}
\end{figure}

Industrial robots are requested to be faster and more accurate. We aim to elaborate efficient algorithms able to command the actuators of an articulated robot. A jointed arm robot closely resembles the human arm, frequently it is called anthropomorphic arm.  Three basic rotary joints able arm swap, shoulder swivel and elbow rotation. Additional 3 revolute joints (roll, yaw, pitch) allow the robot to point in all directions. The joints are arranged in a chain so that one joint supports another further in the chain. The figure \ref{fig:kukaLBR4} illustrates a $n\,(=7)$ degrees of freedom articulated industrial robot. This type of robotic manipulator is powered by its $n$ electric embedded motors activating the $n$ rotation parameters $q^i$.

\section{Robot Dynamics}

\subsection{Kinetic energy}
The velocities field of the robotic articulated system is a linear function of the time derivatives $\dot q^i$ of the configuration parameters $q^i$. The kinetic energy $W(q,\dot q)$ is a strictly convex quadratic function of the second variable $\dot q$, with a positive definite Hessian $\mathbf M(q)$ called \emph{mass tensor}, depending on $q$ :
\begin{equation}
\label{eq:formeQuadratiqueEnergieCinetique}
	W(q,\dot q) = \frac 12 \sum\limits_{i=1}^n \sum\limits_{j=1}^n \mathbf M_{ij}(q)\dot q^i \dot q^j
\end{equation}

\subsection{Gravitational potential energy}

	The gravity actions are modeled by a potential energy $V(q)$ which is the product of the total mass $m$, the local intensity of the gravitational field, and the height of the mass center. 
	
\subsection{Actuators}
	The $i^{\mathrm{\,th}}$ motor exerts between the $i^{\mathrm{\,th}}$ link and the preceding in the chain a torque of intensity $u_i$. For any virtual variations $\delta q^i$ of the configuration parameters $q^i$, the virtual work of the actuators is $\sum\limits_{i=1}^n u_i\, \delta q^i$. This virtual work being coordinates free, $u_i$ are the covariant components of a tensor that we will call \emph{torque tensor} and denote $u$.

\section{Motion Equations}

	The motion of the system is therefore governed by the $n$ Lagrange's equations : 
\begin{equation}
\label{eq:lagrangesEquations}
	\frac {\mathrm d}{\mathrm dt} \left(\frac{\partial W}{\partial \dot q^i}\right) - \frac{\partial W}{\partial q^i} = u_i - \frac{\partial V}{\partial q^i}
\end{equation}

These motion equations can be itemized as

\[
\sum\limits_{j=1}^{N}\frac{\partial^2 W}{\partial \dot q^i\partial \dot q^j}\ddot q^j + \sum\limits_{l=1}^{N}\frac{\partial^2 W}{\partial \dot q^i\partial q^l}\dot q^l - \frac{\partial W}{\partial q^i} + \frac{\partial V}{\partial q^i} = u_i
\]

The quadratic shape (\ref{eq:formeQuadratiqueEnergieCinetique}) of the kinetic energy function leads to the explicit expression 

\[
\sum\limits_{j=1}^{n}\mathbf M_{ij}\ddot q^j + \sum\limits_{k=1}^{n}\sum\limits_{l=1}^{n}\Gamma_{ikl}\dot q^k\dot q^l + \frac{\partial V}{\partial q^i} = u_i
\]

\noindent where the coefficients $\Gamma_{ikl}=\frac12\left(\frac{\partial \mathbf M_{ik}}{\partial q^l}+\frac{\partial \mathbf M_{il}}{\partial q^k}-\frac{\partial \mathbf M_{kl}}{\partial q^i}\right)$ are the Christoffel symbols (of the first kind) associated to the $\mathbf M_{ij}$ regarded as the coefficients of the Riemannian metric.\smallskip

	Introducing the coefficients $\mathbf M^{ij}$ of the inverse metric tensor and the Christoffel symbols of the second kind $\Gamma_{kl}^j = \mathbf M^{ji}\Gamma_{ikl}$, the Lagrange's equations (\ref{eq:lagrangesEquations}) transform into the ordinary differential equation (ODE) 
\begin{equation}
\label{eq:equationsMouvementUbas}
	\sum\limits_{j=1}^{n} \mathbf M_{ij}\left(\ddot q^j + \sum\limits_{k=1}^{n}\sum\limits_{l=1}^{n}\Gamma_{kl}^j\dot q^k \dot q^l + \sum\limits_{k=1}^{n}\mathbf M^{jk} \frac{\partial V}{\partial q^k}\right) = u_i
\end{equation}
\noindent or equivalently
\begin{equation}
\label{eq:equationsMouvementUhaut}
	\ddot q^i + \sum\limits_{j=1}^{n}\sum\limits_{k=1}^{n}\Gamma_{jk}^i\dot q^j \dot q^k + \sum\limits_{k=1}^{n}\mathbf M^{ik} \frac{\partial V}{\partial q^k} = u^i
\end{equation}

	\textbf{Remark 1:} In the right-hand side of equation (\ref{eq:equationsMouvementUhaut}), following the rules of tensorial calculus in a Riemannian manifold, we have introduced the $i^{\,\mathrm{th}}$ contravariant component $u^i=\sum\limits_{k=1}^{n}\mathbf M^{ik}u_k$ of the torque tensor $u$.\smallskip

	\textbf{Remark 2:} In the left-hand side of equation (\ref{eq:equationsMouvementUhaut}), $\frac{\partial V}{\partial q^k}$ is the $k^{\mathrm{th}}$ covariant component of a tensor and $\sum\limits_{k=1}^{n}{\mathbf M^{ik}}\frac{\partial V}{\partial q^k}$ is its $i^{\mathrm{th}}$ contravariant component.\smallskip
	
	\textbf{Remark 3:} In the left-hand side of equation (\ref{eq:equationsMouvementUhaut}), $\ddot q^i + \sum\limits_{j=1}^{n}\sum\limits_{k=1}^{n}\Gamma^i_{jk}\,\dot q^j\dot q^k$ is the covariant time derivative of $\dot q^i$, it is the $i^{\mathrm{th}}$ contravariant component of a tensor that we will call \emph{acceleration vector} and denoted $\frac{\mathrm{\hat d} \dot q}{\mathrm dt}$; with these notations, equation (\ref{eq:equationsMouvementUbas}) reads $\mathbf M_{ij}\, \frac{\mathrm{\hat d}\dot q^j}{\mathrm dt} = u_i - \frac{\partial V}{\partial q^i}$ according to Newton's prescription.

\section{Time Discretization}


To solve the motion equations (\ref{eq:equationsMouvementUhaut}), we perform a time discretization based on Hermite Finite Elements technique \cite{Allaire09,Marchouk85,Strang73}.

\subsection{Cubic Hermite Finite Elements}

\subsubsection{Cubic Hermite Functions}
\begin{figure}[htbp]
	\centering
	\begin{tabular}{ccc}
		\vspace{3mm}
		\def\svgwidth{3cm}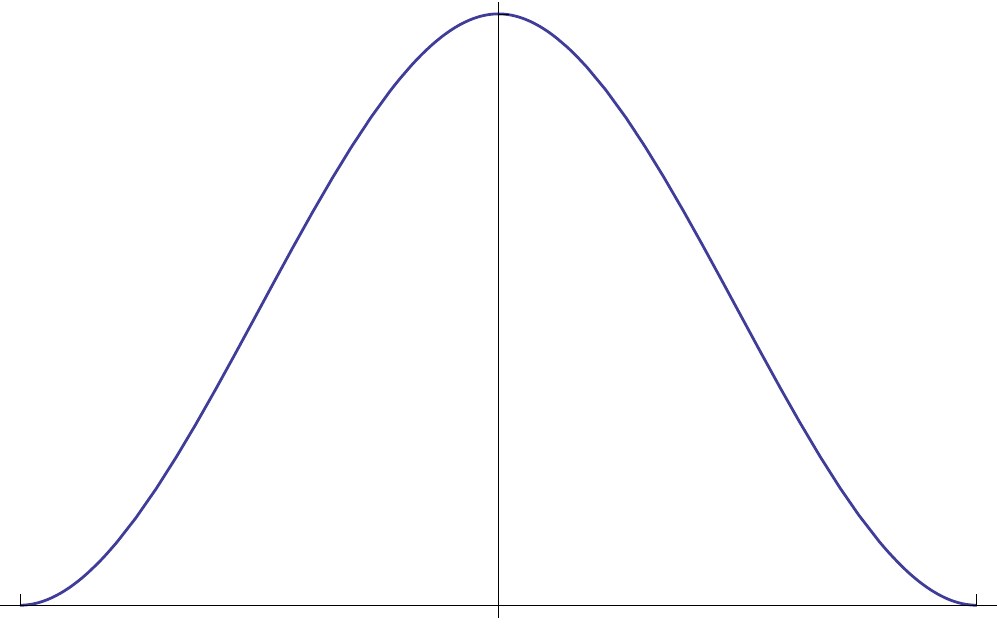&\def\svgwidth{3cm}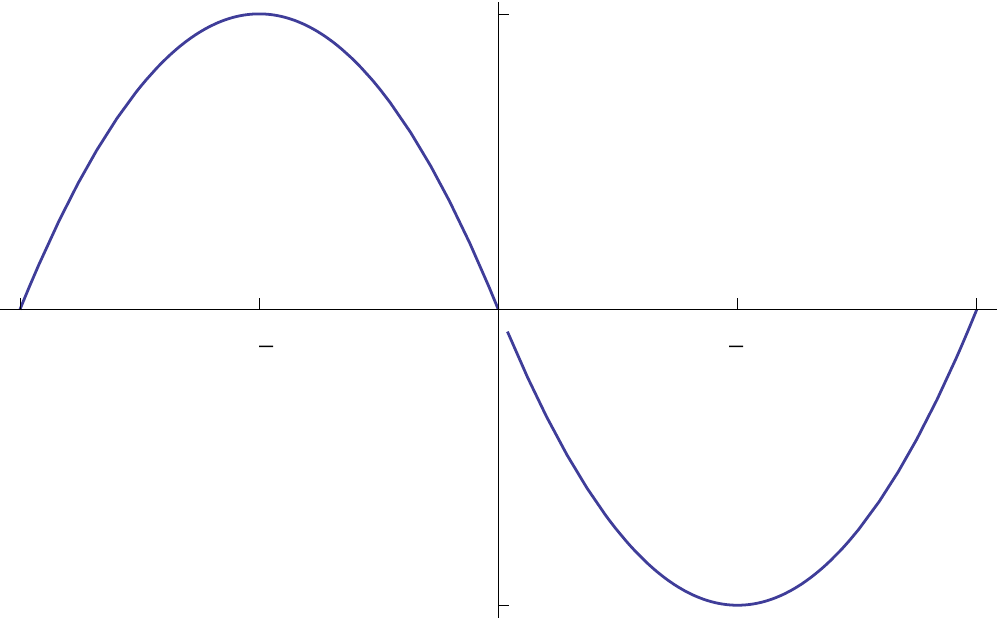&\def\svgwidth{3cm}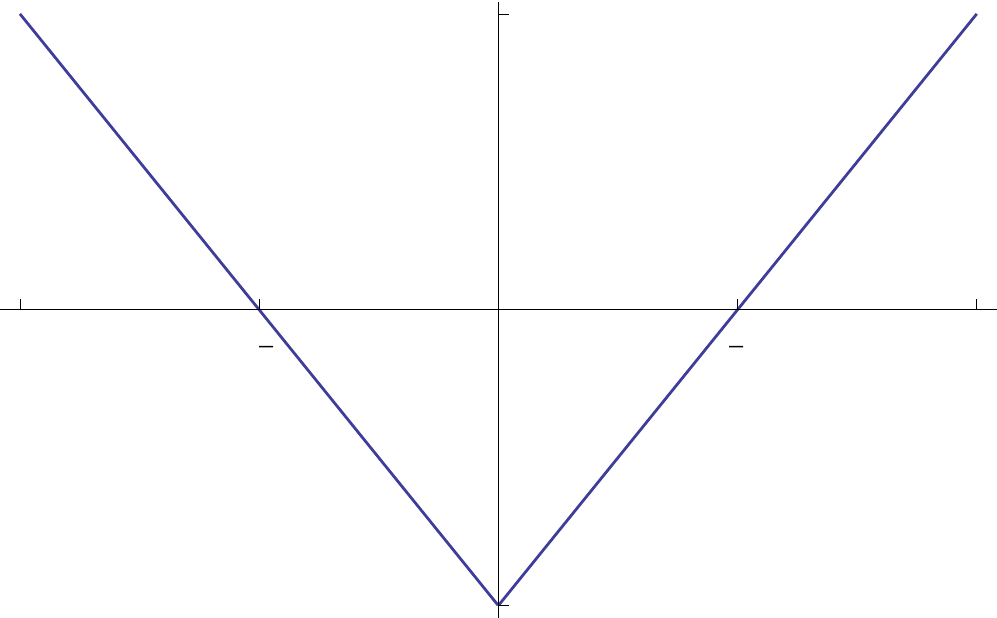\\
		\def\svgwidth{3cm}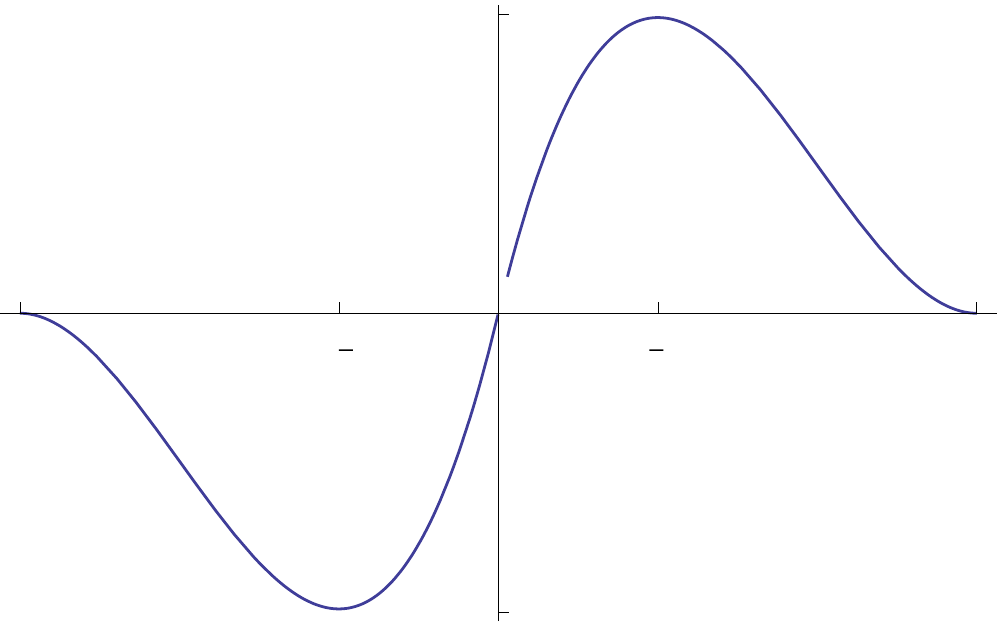&\def\svgwidth{3cm}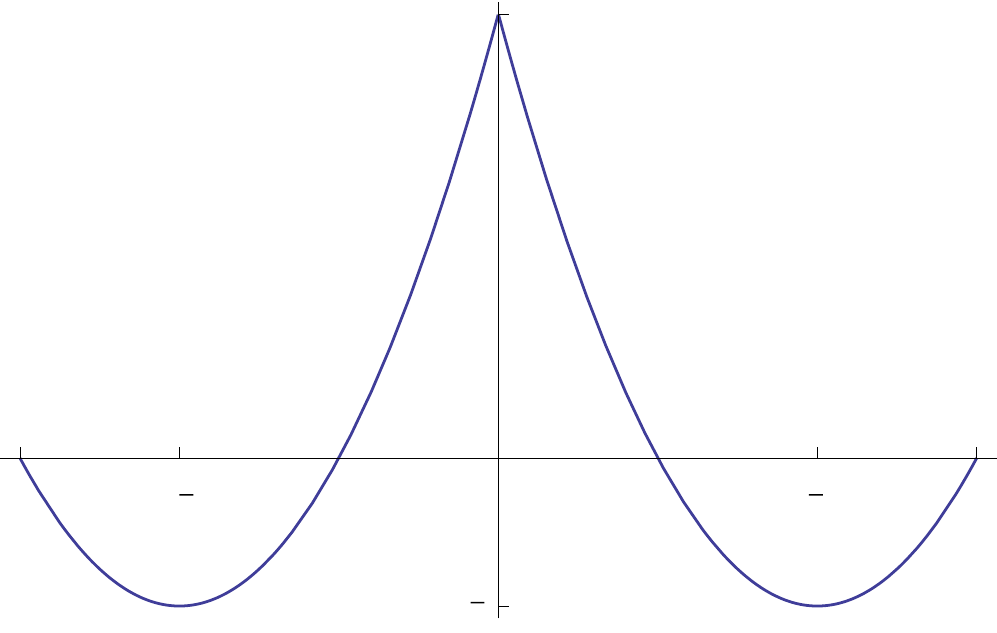&\def\svgwidth{3cm}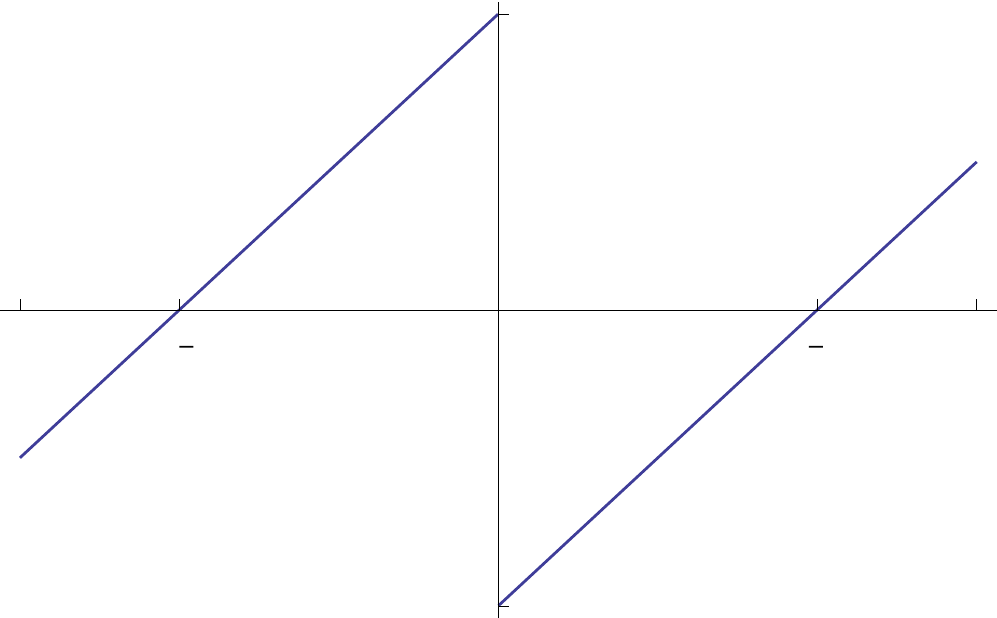
	\end{tabular}
\caption{Hermite Cubics}
\label{fig:hermiteCubics}
\end{figure}

The Cubic Hermite Elements are based on two functions $\phi$ and $\psi$ defined for $t$ belonging to the interval $[-1,+1]$ by the formulae
\[
	\phi(t) = \left(1-\left|t\right|^2\right)\left(1+2\left|t\right|\right) \quad \text{and} \quad \psi(t) = t\left(1-\left|t\right|^2\right)
\]
\noindent and equal to zero outside this interval (see the graphs on Figure \ref{fig:hermiteCubics}).

\subsubsection{Cubic Hermite Finite Elements}
Let us divide the interval $[0,T]$ into $N$ equal pieces of duration $h=T/N$ by the instants $t_p = p\,h\ (p=0\ \text{to}\ N)$. We define the $2N+2$ basis functions 
\[
\phi_p(t) = \phi\left(\frac th - p\right)\quad  \text{and}\quad  \psi_p(t) = \psi\left(\frac th - p\right)
\]

\textbf{Remark:} Each basis function must be truncated when it overflows the interval $[0,T]$ on the left or on the right. This remark is specially important for indices $p=0$ and $p=N$.\smallskip

\subsubsection{Piecewise Cubic Hermite Finite Elements Interpolation}
\label{subsubsec:cubicFEinterpolation}

Each configuration parameter $q^i$ is approximated \cite{Allaire09} by the piecewise cubic Hermite Finite Elements interpolation : 
\begin{equation}
\label{eq:approximationHermiteC}
	q_h^i = \sum\limits_{p=0}^{N}\left(a_p^i\phi_p(t) + h\, b_p^i\psi_p(t)\right)
\end{equation}

The coefficients $a^i_p$ and $b^i_p$ are directly interpreted as the values at time $t=p\,h$ of parameters $q^i$ and derivatives $\dot q^i$.

\subsection{Quintic Hermite Finite Elements}

\subsubsection{Quintic Hermite Functions}
\begin{figure}[htbp]
	\centering
	\begin{tabular}{ccc}
		\vspace{3mm}
		\def\svgwidth{3cm}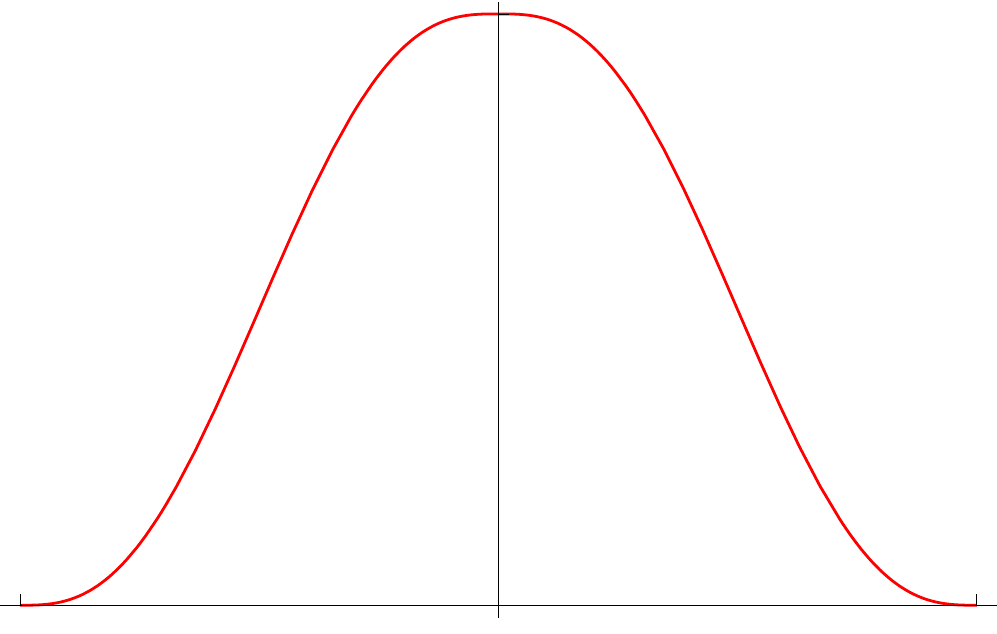&\def\svgwidth{3cm}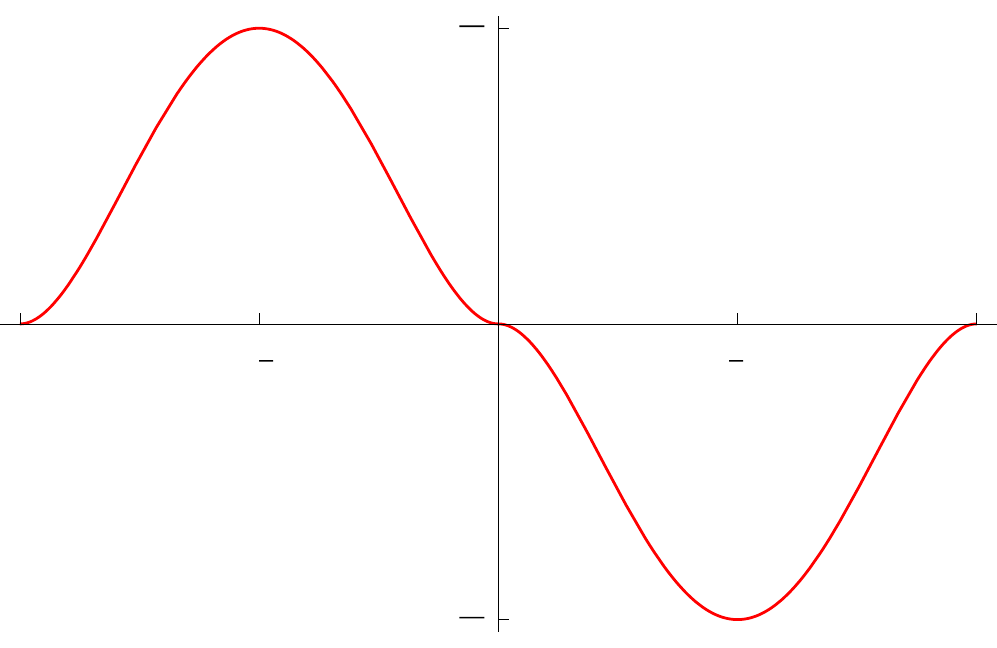&\def\svgwidth{3cm}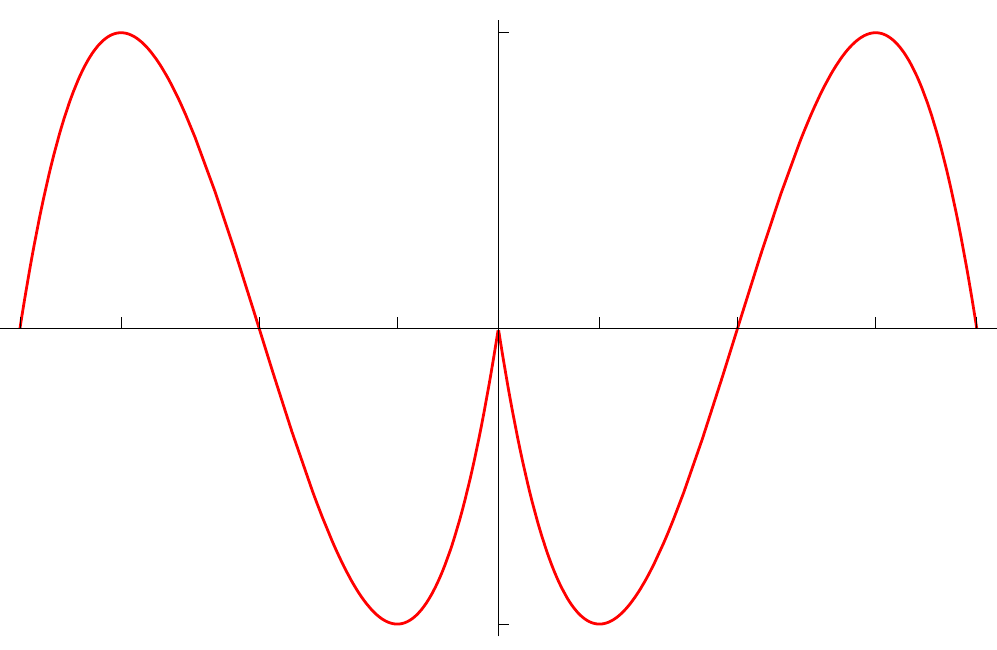\\
		\vspace{3mm}
		\def\svgwidth{3cm}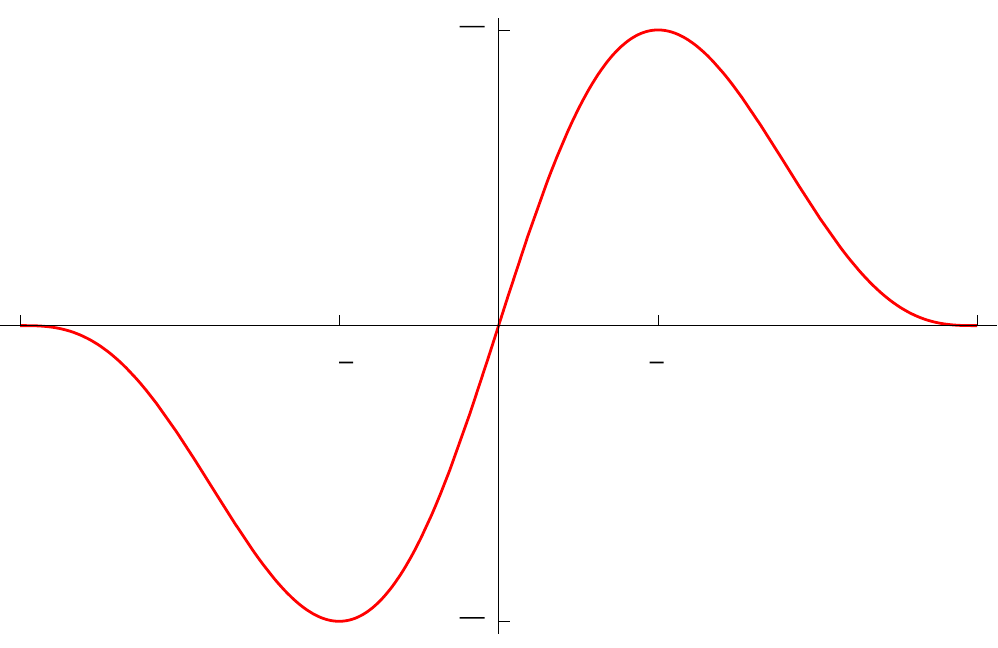&\def\svgwidth{3cm}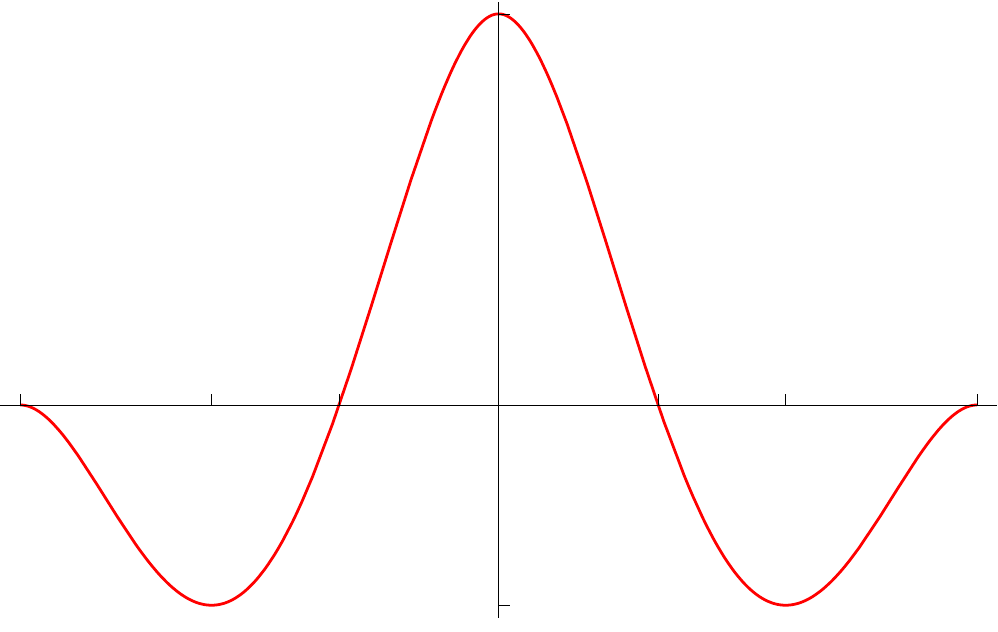&\def\svgwidth{3cm}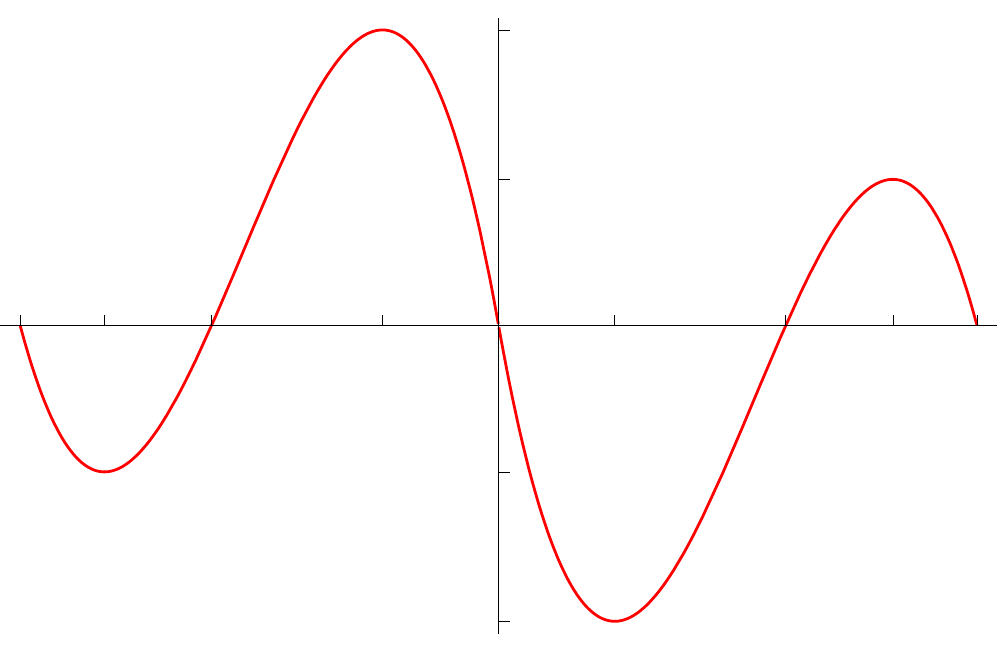\\
		\def\svgwidth{3cm}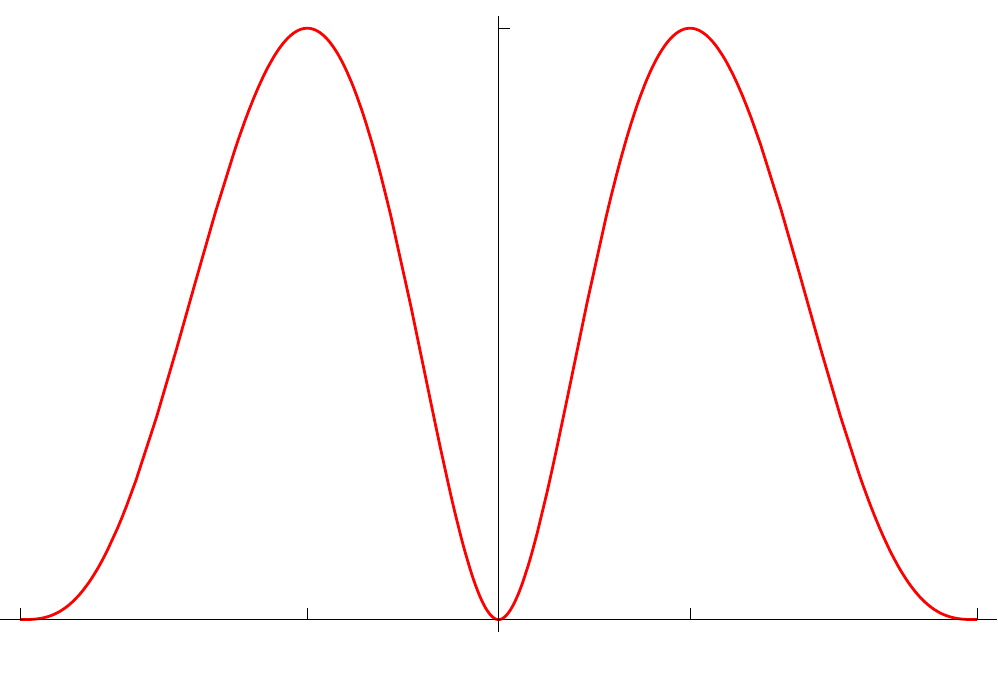&\def\svgwidth{3cm}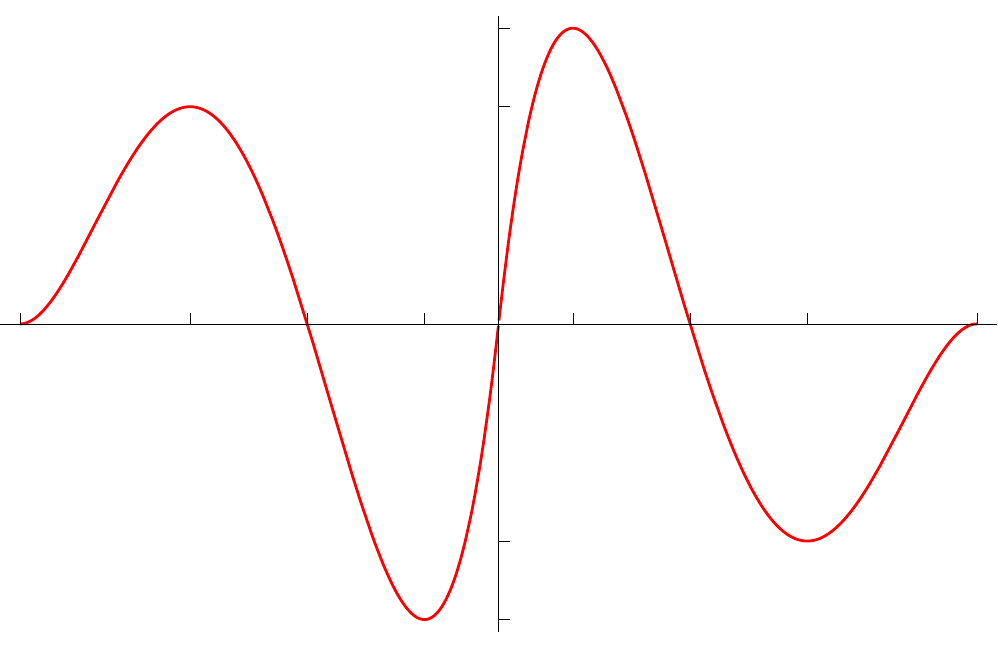&\def\svgwidth{3cm}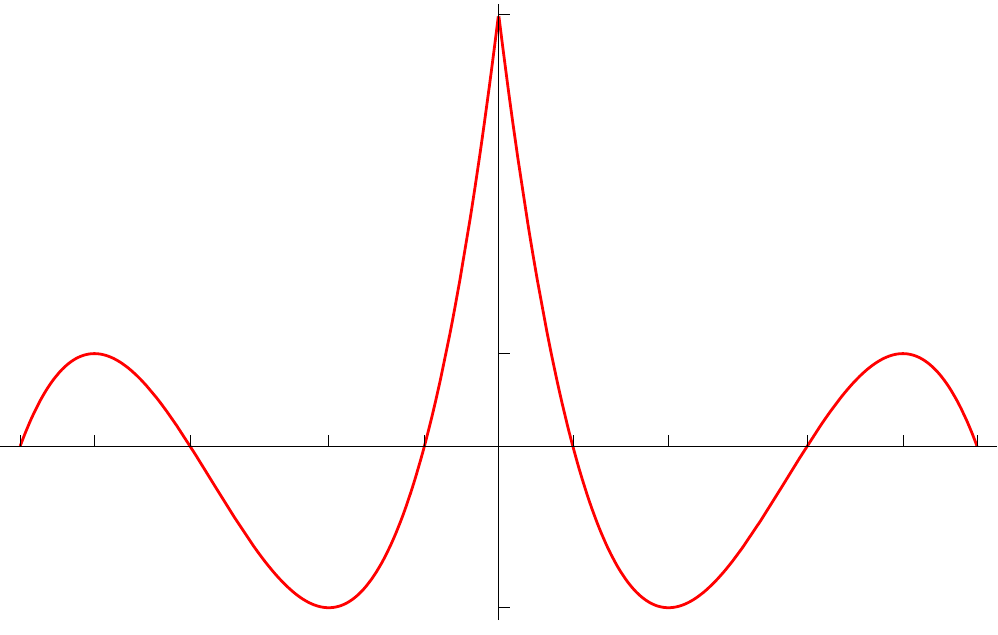
	\end{tabular}
\caption{Hermite Quintics}
\label{fig:hermiteQuintics}
\end{figure}

The quintic Hermite Finite Elements are based on 3 functions $\phi$, $\psi$ and $\theta$, defined for $t$ belonging to the interval $[-1,+1]$ by the formulae 
\[
	\phi(t) = \left(1-\left|t\right|^3\right)\left(1+3\left|t\right| + 6t^2\right) ,\quad \psi(t) = t\left(1-\left|t\right|\right)^3 \left(1+3\left|t\right|\right) ,\quad \theta(t) =	\frac{t^2}2\left(1-\left|t\right|\right)^3 
\]
\noindent and equal to zero outside (see the graphs on Figure \ref{fig:hermiteQuintics}).

\subsubsection{Quintic Hermite Finite Elements}
Let us divide the interval $[0,T]$ into $N$ equal pieces of duration $h=T/N$ by the instants $t_p = p\,h\ (p=0\ \text{to}\ N)$. We define the $3N+3$ basis functions 
\[
\phi_p(t) = \phi\left(\frac th - p\right), \quad \psi_p(t) = \psi\left(\frac th - p\right)\quad \text{and} \quad \theta_p(t) = \phi\left(\frac th - p\right)
\]

\subsubsection{Piecewise Quintic Hermite Finite Elements Interpolation}
\label{subsubsec:quinticFEinterpolation}

Each configuration parameter $q^i$ is approximated \cite{Allaire09} by the piecewise cubic Hermite Finite Elements interpolation : 
\begin{equation}
\label{eq:approximationHermiteQ}
	q_h^i = \sum\limits_{p=0}^{N}\left(a_p^i\phi_p(t) + h\, b_p^i\psi_p(t) + h^2\, c_p^i \theta_p(t)\right)
\end{equation}

The coefficients $a^i_p$, $b^i_p$ and $c^i_p$ are directly interpreted as the values at time $t=p\,h$ of parameters $q^i$ and derivatives $(\dot q^i, \ddot q^i)$.

\subsection{Accuracy of the interpolations}
In one dimension, let us consider for example, the circular function $q(t)=\sin t$. Then the norm (respectively $\|e\|_5 = \left(\int_0^T\left[(e(t))^2 + h^2(\dot e(t))^2 + h^4(\ddot e(t))^2 \right]\dt \right)^{\frac12}$) of the error $e(t)=q_h(t)-\sin t$ reveals to be in $h^4$ (respectively in $h^6$) for the cubic (respectively quintic) Hermite Finite Elements interpolation.

\section{Motion Simulations}

\subsection{Time integration algorithm}

When the history $u(t)$ of the torque tensor is known, the state vector $x(t)=\begin{bmatrix} q(t)\\ \dot q(t) \end{bmatrix}$ of the robot can be predicted at any time $t$ from its initial value $x(0)=\begin{bmatrix} q(0)\\ \dot q(0) \end{bmatrix}$ by solving numerically \cite{Siebert12} the ODE (\ref{eq:equationsMouvementUhaut}). To simulate the trajectories of the robot, we design the following integration algorithm :\smallskip

\begin{enumerate}[(i)]
	\item	approximate each configuration parameter $q^i$ by its piecewise cubic (respectively quintic) Hermite Finite Elements interpolation as in formula (\ref{eq:approximationHermiteC}) (respectively (\ref{eq:approximationHermiteQ})),
	\item	express the motion equations (\ref{eq:equationsMouvementUhaut}) at $2nN$ (respectively $3n(N+1)$) well suited instants,
	\item	solve this algebraic system concerning the $2nN$ (respectively $3n(N+1)$) unknown coefficients $(a_p^i, b_p^i)$ (respectively $(a_p^i, b_p^i,c_p^i)$),
	\item	rebuild the approximation $q^i_h$.
\end{enumerate}\smallskip

\textbf{Remark 1:} The steps (ii) and (iii) of the above algorithm are inspired from the inertial parameters identification technique applied in \cite{Gautier92}.\smallskip

\textbf{Remark 2:} The coefficients $a^i_0$ and $b_0^i$ are known from the initial conditions $(a^i_0 = q^i(0), b^i_0 = \dot q^i(0))$.

%
\subsection{Example}

In one dimension, let us consider the nonlinear pendulum equation $\matrice M \ddot q + mgl\sin q = 0$\ \ with the initial conditions $q(0)=\frac{\pi}{12}$ and $\dot q(0)=0$. The inertia coefficient $\matrice M$, mass $m$ and the length $l$ are such that $\omega = \sqrt{\frac gl} = 3.102\, s^{-1}$. Implementing the above Time Integration Algorithm for cubic Hermite Finite Elements with $N=35$, we found a \emph{periodic} solution $q_h$. The identified period of oscillation $\tau=2.03452\, s$ coincides up to \emph{machine precision} to the period provided by calculating the Legendre's elliptic integral of the second kind $\displaystyle \frac4\omega \int_0^{\frac\pi2}\left(1-\left(\frac{\pi}{24}\right)^2\sin^2\alpha\right)^{-\frac12}\, \mathrm d\alpha$.
\section{Optimal Control}
\label{sec:optimalControl}

Initially, the robotic articulated system is in a state of positions and velocities $x_0 = x(0)$. In a fixed final time $T$, we want to bring it  to a final state $x_1 = x(T)$. What is the torque tensor needed to perform this task? The answer to this question is not unique. In order to bound their intensities, the torques are selected by minimizing an integral functional $J(u)=\int\limits_{0}^{T} \gamma\left(u(t)\right)\, \mathrm dt$ \ called objective functional. The integrand $\gamma$ is a convex function called cost function.

\section{Invariant Cost Function}

Usually, the cost function is chosen as a quadratic mean of the covariant components $u_i$ of the torque tensor	$\gamma (u) = \frac 12 \sum\limits_{i=1}^{n}\sum\limits_{j=1}^{n} \mathbf S^{ij}u_i\, u_j$, where $\mathbf S$ is a symmetrical positive definite bilinear form focusing on main torques. We will choose the tensor $\mathbf S$ so that $\mathbf S^{ij} = \mathbf M^{ij}$. With this choice, the cost function $\gamma(u)=\frac12 \sum\limits_{i=1}^n u_i\,u^i$ is coordinates free.

\section{Optimization Method}

The original question asked in paragraph \ref{sec:optimalControl} enters within the frame of the classical calculus of variations : Minimize the integral functional
\begin{equation}
\label{eq:critereI(q)}
	I(q) = \int\limits_0^T \Lagr\left(q(t),\dot q(t), \ddot q(t)\right)\,\mathrm dt
\end{equation}
\noindent where the integrand is the Lagrangian function 
\[
	\Lagr\left(q,\dot q, \ddot q\right) = \frac12 \sum\limits_{i=1}^{n}\sum\limits_{j=1}^{n}\matrice M_{ij} u^i u^j
\]
\noindent with $u^i$ and $u^j$ expressed in terms of $\left(q,\dot q, \ddot q\right)$ by equation (\ref{eq:equationsMouvementUhaut}). The Euler-Lagrange equations reads:
\begin{equation}
\label{eq:equationsMvtLagrange}
\frac{\ddroit ^2}{\ddroit t^2}\left(\frac{\partial \Lagr}{\partial \ddot q^i}\right) - \frac{\ddroit}{\ddroit t}\left(\frac{\partial \Lagr}{\partial \dot q^i}\right) + \frac{\partial \Lagr}{\partial q^i} = 0
\end{equation}

\noindent Because of our choice of an invariant cost function $\gamma$, these equations will reveal to be covariant as advocated by Einstein for the modeling of any physical phenomena. We can remark that

\[
\begin{array}{ccc}
\displaystyle
\frac{\partial \Lagr}{\partial \ddot q^i}=u_i	\quad\text{and}\quad
&\displaystyle
\frac{\partial \Lagr}{\partial \dot q^i}=2\sum\limits_{j=1}^{n}\sum\limits_{k=1}^{n}\Gamma_{ik}^j u_j \dot q^k
\end{array}
\]

Therefore, equations (\ref{eq:equationsMvtLagrange}) are \second\ order ODE in the dual variables $(q^i,u_i)$. We will call them \emph{control equations}. Associated to the \emph{motion equations} (\ref{eq:equationsMouvementUhaut}), they provide a system of $2n$ second order ODE for finding the torques and the trajectories. This system is an alternative to the Pontryagin's system of $4n$ first order ODE. But, with our point of view, the adjoint parameters are the torques $u_i$ which are directly interpretable.

\section{Optimal Control Algorithm}


Insertion of the quintic approximation (\ref{eq:approximationHermiteQ}) of $q$ in $I(q)$ generates a function $I(q_h)$ depending solely on the coefficients $a_p^i$, $b_p^i$ and $c_p^i$. These coefficients are obtained, in finite dimension, by minimizing $I(q_h)$ with the Polak-Ribière's conjugate gradient method \cite{Joly86, Polak71, Polak69}.

\textbf{Remark}: The coefficients $a_p^i$ and $b_p^i$ are known for $p=0$ and $p=N$
\[
a_0^i = q^i(0),\quad b_0^i=\dot q^i(0),\quad a_N^i=q^i(T),\ \text{and}\quad  b_N^i = \dot q^i (T)
\]

After obtaining an approximation of $q(t)$, we obtain an approximation of the optimal torque $u(t)$ by coming back to equation (\ref{eq:equationsMouvementUbas}).
\section{Validation of the Optimal Control Algorithm}

As an example, we consider a simple mechanical system with one degree of freedom. It is governed by the motion equation $\ddot q + q = u$. The final time $T$ is equal to $\pi$. The initial conditions are $q(0)=0$ and $\dot q(0) =1$. The final conditions are $q(\pi)=0$ and $\dot q(\pi) =-1$. The objective function is $J(u) = \frac 12 \int\limits_0^T (u(t))^2\,\mathrm dt$. The \emph{control equation} (\ref{eq:equationsMvtLagrange}) is reduced to $\ddot u + u = 0$. The optimization method summarizes in : Minimize $I(q)=\frac 12 \int\limits_0^T\left(q(t)+\ddot q(t)\right)^2\,\mathrm dt$. The theoritical solution is $q(t)=\sin t$, $\dot q(t) = \cos t$ and $u(t)=0$ for which the minimum value $I(q)=0$ is achieved. Implementing the above Optimal Control Algorithm, the positive sequence $I(q_h)$ reveals to be decreasing as $h^6$. The norm $\|\, \|_5$ of the error $q-q_h$ also reveals to decrease as $h^6$, confirming the superconvergence \cite{Allaire09} of the Hermite's technique.



\section{Outlook}

We plan to apply our approach in the context of grasping in real time a falling ball \cite{Maire01} with the last manipulator link of an articulated robot. The extension to the interception of a moving ball in a free final time will follow the method we exhibited in a previous work \cite{Lazrak95}.

\section{Conclusions}

The robot dynamics is modeled according to Lagrange's analytical mechanics, with the same geometrical requirements that general relativity \cite{Misner73}. As an alternative to Pontryagin's Maximum Principle, our optimal control algorithm is developed in the frame of the classical Lagrange's calculus of variations. We have presented a piecewise quintic Hermite Finite Elements Method for computing an accurate approximation of the optimal trajectories and controls. These Hermite Elements sound to be well suited for generating smooth, fast motion for a mobile robot in a changing environment, with good efficiency and stability.



\end{document}